\documentclass[conference]{IEEEtran}
\IEEEoverridecommandlockouts
\usepackage[printonlyused]{acronym}
\usepackage{cite}
\usepackage{amsmath,amssymb,amsfonts}
\usepackage[inkscapeformat=png]{svg}
\usepackage{svg}
\usepackage{graphicx}
\usepackage{tikz}
\usepackage{pgfplots}
\usepackage{pgfplotstable}
\usepackage{ifthen}
\pgfplotsset{compat=1.7}
\usepackage{soul}
\usepackage[font=footnotesize]{subcaption}
\usepackage{tabularx}
\usepackage[font=footnotesize]{caption}
\usepackage{booktabs}
\usepackage{bbm}

\usepackage{textcomp}

\usepackage{amsmath}
\DeclareMathOperator*{\argmax}{arg\,max}
\DeclareMathOperator*{\argmin}{arg\,min}
\usepackage{amssymb}
\usepackage{optidef}
\usepackage{algorithm}
\usepackage{enumitem}
\usepackage{hyperref}
\usepackage{balance}
\usepackage{xcolor}
\usepackage{mathtools}

\usepackage{algpseudocode}

\newcommand*\mb[1]{\mathbf{#1}}

\pgfplotsset{every axis/.append style={
		scaled x ticks = false,
		label style={font=\footnotesize},
		tick label style={font=\footnotesize},
		tick scale binop=\times
	}
}

\def\BibTeX{{\rm B\kern-.05em{\sc i\kern-.025em b}\kern-.08em
    T\kern-.1667em\lower.7ex\hbox{E}\kern-.125emX}}

\title{BS-Breath: Respiration Sensing with Cell-free Massive MIMO}

\author{
\IEEEauthorblockN{
Haoqiu Xiong\IEEEauthorrefmark{1}, Robbert Beerten\IEEEauthorrefmark{1}, Zhuangzhuang Cui\IEEEauthorrefmark{1}, Yang Miao\IEEEauthorrefmark{2}, Sofie Pollin\IEEEauthorrefmark{1}
}
\IEEEauthorblockA{
\IEEEauthorrefmark{1}\textit{WaveCoRE, Departement of Electrical Engineering (ESAT), KU Leuven, Belgium} \\
\textit{\IEEEauthorrefmark{2} EEMCS faculty, University of Twente, The Netherlands}
}%
\thanks{This work is supported by the Horizon Europe Research and Innovation programme with Grant Agreement No. 101192521 (MultiX), and No. 101096954 (6G-BRICKS). The code and the data for the experiments can be found at: https://gitlab.kuleuven.be/u0149002/bs-breath}
}
\begin{document}
\maketitle%
\begin{abstract}
This paper demonstrates the feasibility of respiration pattern estimation utilizing a communication-centric cell-free massive MIMO OFDM \ac{BS}. The sensing target is typically positioned near the \ac{UE}, which transmits uplink pilots to the \ac{BS}.
Our results demonstrate the potential of massive MIMO systems for accurate and reliable vital sign estimation. 
Initially, we adopt a single antenna sensing solution that combines multiple subcarriers and a breathing projection to align the 2D complex breathing pattern to a single displacement dimension. 
Then, \ac{WAC} aggregates the 1D breathing signals from multiple antennas. The results demonstrate that the combination of space-frequency resources—specifically in terms of subcarriers and antennas—yields higher accuracy than using only a single antenna or subcarrier. Our results significantly improved respiration estimation accuracy by using multiple subcarriers and antennas. With \ac{WAC}, we achieved an average correlation of 0.8 with ground truth data, compared to 0.6 for single antenna or subcarrier methods—a 0.2 correlation increase. Moreover, the system produced perfect breathing rate estimates. These findings suggest that the limited bandwidth (18~MHz in the testbed) can be effectively compensated by utilizing spatial resources, such as distributed antennas. 

\end{abstract}
\acrodef{ISAC}{Integrated Sensing and Communications}
\acrodef{MIMO}{Multiple-Input and Multiple-Output} 
\acrodef{BS}{Base Station}
\acrodef{UL}{Uplink}
\acrodef{UE}{User Equipment}
\acrodef{LTE}{Long-Term Evolution}
\acrodef{TDD}{Time Division Duplexing}
\acrodef{TDD LTE}{Time Division Duplexing-Long Term Evolution}
\acrodef{CSI}{Channel State Information}
\acrodef{CIR}{Channel Impulse Response}
\acrodef{CFR}{Channel Frequency Response}
\acrodef{OFDM}{Orthogonal Frequency-Division Multiplexing}
\acrodef{IFFT}{Inverse Fast Fourier Transform}
\acrodef{LoS}{Line-of-Sight}
\acrodef{NLoS}{Non-Line-of-Sight}
\acrodef{SA}{Subcarrier Aggregation}
\acrodef{MRC}{Maximum Ratio Combining}
\acrodef{MAC}{Multiple Antenna Combining}
\acrodef{WAC}{Weighted Antenna Combining}
\acrodef{SNR}{Signal-to-Noise Ratio}
\acrodef{BNR}{Breathing-to-Noise Ratio}
\acrodef{AoA}{Angle-of-Arrival}
\acrodef{IDFT}{Inverse Discrete Fourier transform }
\acrodef{Mocap}{Motion Capture System}

\acrodef{GT}{Groud Truth}
\acrodef{SDR}{Software-Defined Radio}
\acrodef{RF}{Radio Frequency}
\acrodef{bpm}{Breaths Per Minute}

\section{Introduction}

In recent years, the increase of wireless devices has not only strained the available spectrum resources but also highlighted the need for innovative technological solutions that can multitask, fulfilling the dual roles of communication and sensing. Integrated Sensing and Communications (ISAC) represents a pivotal advancement in this direction, promising enhanced spectrum efficiency and resource reusability\cite{isacFan}. This paper delves into a novel ISAC application: utilizing a Base Station (BS) infrastructure for passive respiration monitoring, an essential component of health monitoring systems. By re-purposing existing massive MIMO-OFDM communication systems, we propose a cost-effective deployment strategy that does not demand additional spectrum resources or an infrastructure overhaul.

The advent of massive MIMO systems has revolutionized communications by significantly improving signal quality and channel capacity \cite{mimoofdm_CE,mimo_antenna}. Additionally, in sensing, research \cite{NF_MIMO} has demonstrated that near-field massive MIMO can compensate for limited bandwidth while achieving high-range resolution at the centimeter level.
However, its potential in vital signs sensing remains largely untapped. Previous studies \cite{MIMO_radar_vital_1, MIMO_radar_vital_2} have explored beamforming techniques for multiperson vital signs sensing with MIMO radar, yet they did not investigate the spatial diversity of MIMO for enhancing sensing accuracy. Research like DiverSense \cite{DiverSense} extended the Wi-Fi sensing range by leveraging subcarrier and antenna diversity. However, typical Wi-Fi systems have limited antennas (e.g., up to three antennas with an Intel 5300 Wi-Fi chip\cite{intel5300}). This motivates our investigation in this paper. Specifically, we investigate how multiple subcarriers and antennas can be combined in the massive MIMO-OFDM system to enhance the accuracy and reliability of respiration estimation, which is crucial in health monitoring applications.

\textbf{Contributions:} To the best of our knowledge, we are the first to demonstrate the practical use of a cell-free massive MIMO \ac{BS} for highly precise vital signs sensing. Specifically, our study evaluates the performance improvements obtained through multiple subcarrier and antenna signal processing techniques such as \ac{IDFT}\cite{ofdmradar,decimeterWiFi}, DiverSense\cite{DiverSense}, and the \ac{WAC} across multiple antennas. Unlike other approaches, our method does not rely on specific UE or BS orientation relative to the user, thanks to the spatial diversity enabled by cell-free MIMO.


\section{System Model}
\begin{figure}[t]
    \centering
    \centerline{\includegraphics[width=8.5cm]{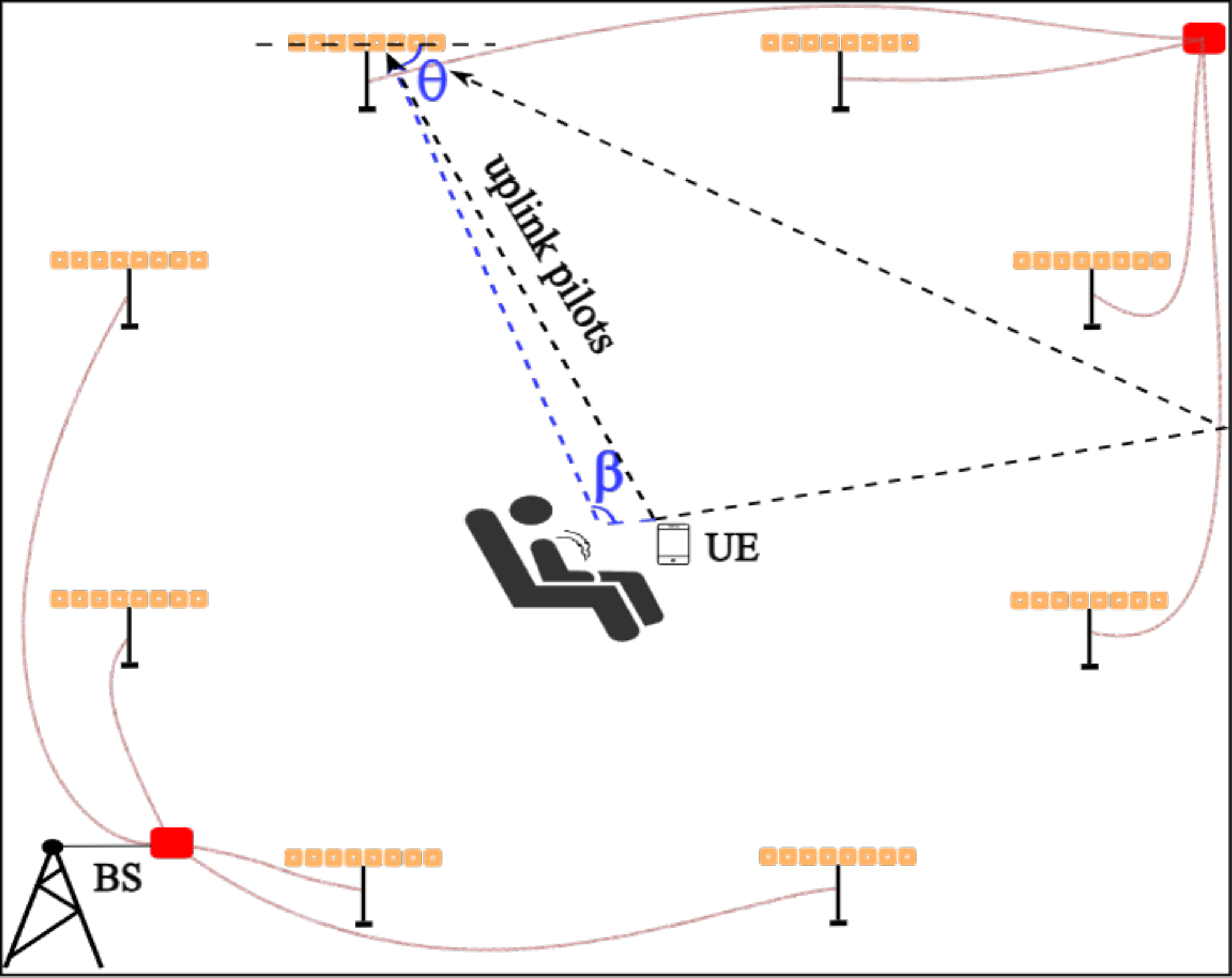}}
  \caption{Cell-free uplink respiration sensing.}
  \label{fig:uplinksensing}
\end{figure}

Our analysis considers a cell-free Massive MIMO communication system operating under the \ac{TDD} mode, utilizing an OFDM waveform. The \ac{BS} estimates the \ac{CSI} of the \ac{UL} channel over the full bandwidth of the system. This \ac{CSI} is estimated from known orthogonal pilot sequences transmitted by the \ac{UE}s, as shown in Fig. \ref{fig:uplinksensing}.


The \ac{UL} pilots are transmitted at regular intervals $i = 1 \dots I$ of $T_s$ seconds each with $I$ total frames 
per measurement. Each frame consists of subcarriers $k=1 \dots K$, each equally spaced at $\Delta f$ Hz within the total bandwidth $B$. 
Therefore the \ac{CSI} estimated by the $M$ 
antennas of the \ac{BS} can be represented by the complex tensor: 

\begin{equation}
    H_\text{CSI}  \in \mathbb{C}^{M\times K \times I}.
\end{equation}

We consider there are $L$ distinct paths between \ac{UE} and \ac{BS}, each indexed by $l\in  1 \dots L$. Each path is characterized by amplitudes $\alpha_l$ and propagation delays $\tau_l$. The measured \ac{CFR} at the $m$-th antenna of \ac{BS}, on the $k$-th subcarrier and $i$-th frame, is represented as
\begin{equation}
\label{eq:cfr}
        h_{mk}[i] = \sum_{l=1}^{L}  \alpha_{mkl}[i] e^{-j 2 \pi (f_c + k\Delta f  ) \tau_{ml}[i]},
\end{equation}
where $\alpha_{mkl}[i]$ and $\tau_{ml}[i]$ may vary over time, particularly in the presence of moving targets such as a chest. In addition, the multipath also differs across various received antennas 
due to the diverse propagation geometries between the transmitter and receiver. 
The path delay $\tau_{ml}[i]$ can be expanded as
\begin{equation}
\label{eq:delay}
\tau_{ml}[i] =  \frac{\dot{d}_{ml}+b[i]}{c} (m-1)\sin(\theta_{ml}[i]) \cos(\frac{\beta_{ml}[i]}{2}), 
\end{equation}
where $\dot{d}_{ml}$, $b[i]$, $\theta_{ml}[i]$, and $\beta_{ml}[i]$ refer to the initial path length, the chest displacement during the $i$-th frame, the \ac{AoA}, and the bistatic angle \cite{cherniakov2001bistatic}, respectively. $\cos(\frac{\beta_{ml}[i]}{2})$ is the bistatic projection factor from target movement to changes in multipath length\cite{WiGigVital}.


\section{Signal Processing}
The underlying principle of respiration estimation is straightforward: displacement of the chest during breathing alters the lengths of multipath components reflecting off the chest. This change, in turn, affects both the phase and amplitude of the received multipath components, as shown in Eqs. (\ref{eq:cfr}) and (\ref{eq:delay}). Because breathing displacement $b_i$ leads to different phase shifts in the different subcarriers and antennas, the question becomes how to combine these many subcarriers and antennas optimally.
This section compares different signal processing algorithms designed explicitly for respiration sensing.

\subsection{Evaluation Methods}
To evaluate the performance of the proposed methods, we introduce our evaluation method for the estimated breathing signals here. We cross-correlate the estimated power normalized with ground-truth breathing signals obtained using a camera-based \ac{Mocap}. The correlation value ranges between 0 and 1, where 1 means that the estimation is identical to the ground truth. This metric is more informative than the average \ac{bpm}, especially with varying breathing rates. An example is shown in Fig.~\ref{fig:antennadiff}, where the first antenna gives a correlation value of $0.826$ while obtaining the same \ac{bpm} as the ground truth.

\subsection{Multiple Subcarrier Signal Processing} 
The first question is how to combine multiple subcarriers $K$.  Signals reflecting off the chest and surroundings create multiple paths that sum at the receiver, sometimes coherently adding or canceling each other out, affecting the estimated breathing patterns. Two main methods are applied to solve the multipath fading: 1) \ac{IDFT}\cite{ofdmradar} and 2) DiverSense\cite{DiverSense}, and then determine a single estimated range signal $s_m[i]$ per antenna $m$ and frame index $i$. 

\subsubsection{IDFT}
\label{sams}
The IDFT converts the \ac{CFR} into a \ac{CIR} which corresponds to the radar range profile\cite{ofdmradar}. The coefficients of the target delay/range can be estimated by
\begin{equation}
\begin{split}
s_m[i] & = \argmax_{\hat{\tau}_m} \text{IDFT} \big\vert h_{mk}[i] \big\vert \\
       & = \argmax_{\hat{\tau}_m} \big\vert\frac{1}{K}\sum_{k=1}^K h_{mk}[i]e^{j\frac{2\pi}{K}k \hat{\tau}_m} \big\vert, \hat{\tau}_m=1,\dots,K,
\end{split}  
\end{equation}
where $\hat{\tau}_m$ is the estimated delay bin index by $k$. 
Note that the precision of this estimation is constrained by the range resolution $\Delta d = c/B = $16.66 m and the maximum unambiguous range $d_{max} = c/{\Delta f}$=1666 m, where $c$ is the speed of light, $B=\text{18 MHz}$ is the bandwidth, and $\Delta f=\text{180 KHz}$ is the subcarrier spacing. Given these parameters, respiratory movements at the centimeter level consistently occur within the same range of cells. However, due to the coarse range resolution, any path with a difference of less than 16.66 m from the target path will likely be included in the identified range cell, such as the line-of-sight (LoS) path.

\subsubsection{DiverSense}
The initial phase varies in the subcarriers because of the wavelength variation. To coherently combine signals from multiple subcarriers, DiverSense picks one subcarrier signal (e.g., 1-th subcarrier) as the reference $h_{m1}[i]$ and rotates the rest of the signals to align the initial phase with the reference\cite{DiverSense}. The DiverSense integrates multiple subcarriers with a steering vector $e^{-j\gamma_{mk}}$, 
\begin{equation}
    s_m[i] = \frac{1}{K}\sum_{k=1}^{K}h_{mk}[i] e^{-j\gamma_{mk}}.
\end{equation}
The steering vector is estimated by minimizing the distance $D$ between the reference subcarriers and the rest by searching all angles $\gamma_{mk}\in[0,2\pi]$,
\begin{equation}
    \argmin_{\gamma_{mk}} D_k = \frac{1}{I}\sum_{i=1}^{I} \left\vert h_{m1}[i] - h_{mk}[i] e^{-j\gamma_{mk}}\right\vert^2.
\end{equation}




\subsection{Per-Antenna Projection}
After the 
subcarrier signal processing, the selected range 
signal $s_m[i]$ 
can be represented as the complex 
matrix
$S  \in \mathbb{C}^{M \times I}$. 
To deal with the "blind spots" issue studied in \cite{blindspot}, we combine the real part and imaginary part of each antenna signal $s_{m}[i]$ for sensing, which is equivalent to combining amplitude and phase. According to \cite{FarSense}, breathing is a rotation of a complex range signal $s_m[i]$ in a complex space. As breathing corresponds to small movements, only part of a circle is considered during in- or exhaling. Depending on the part of the circle (e.g., top or right), a projection on the real or imaginary axis is most informative. In \cite{FarSense} it is shown that there is an optimal projection axis capturing most of the breathing in a single real estimate, by projecting on an axis that aligns best with the part of the circle in complex space covered by the breathing signal. 
We employ the Farsense algorithm \cite{FarSense} for the individual antennas during preprocessing to achieve a robust estimation. The process initiates by generating a set of combination candidates $ \{ [\cos{\alpha_m}~ \sin{\alpha_m}] : \alpha_m \in [0, 2\pi] \}$, each representing a linear combination of the real ($I_m[i]$) and imaginary ($Q_m[i]$) components of $s_{m}[i]$ :
\begin{equation}
    \hat{s}_m[i] = I_m[i]\cos{\alpha_m} + Q_m[i]\sin{\alpha_m},
\end{equation}
note that $\hat{s}_m[i]$ is a real value after the combination. When $\alpha_m = 0$ for instance, the real part of the complex signal is optimal. 

The method selects the best \ac{BNR} proposed in previous\cite{BNRDina} to evaluate these candidates. The \ac{BNR} is estimated by taking the total breathing power, which is the average power in the frequency band between 0.083 Hz and 0.583 Hz (corresponding to 5 and 35 bpm), and then dividing it by the total power of the preprocessed signal minus the power of the breathing signal, the best BNR in the projection is defined as $\Gamma$: 
\begin{equation}
    \Gamma_m = \frac{P_m^\text{breath}}{P_m^\text{tot} - P_m^\text{breath}}.
\end{equation}
We estimate the final breathing signal $\hat{b}_m[i]$ at each antenna $m$ from the projection $\hat{s}_m[i]$ that maximizes the BNR. 



    


\subsection{Weighted Antenna Combining (WAC)}

\begin{figure}[t]
\begin{minipage}[b]{.49\linewidth}
  \centering
  \centerline{\includegraphics[width=4.8cm]{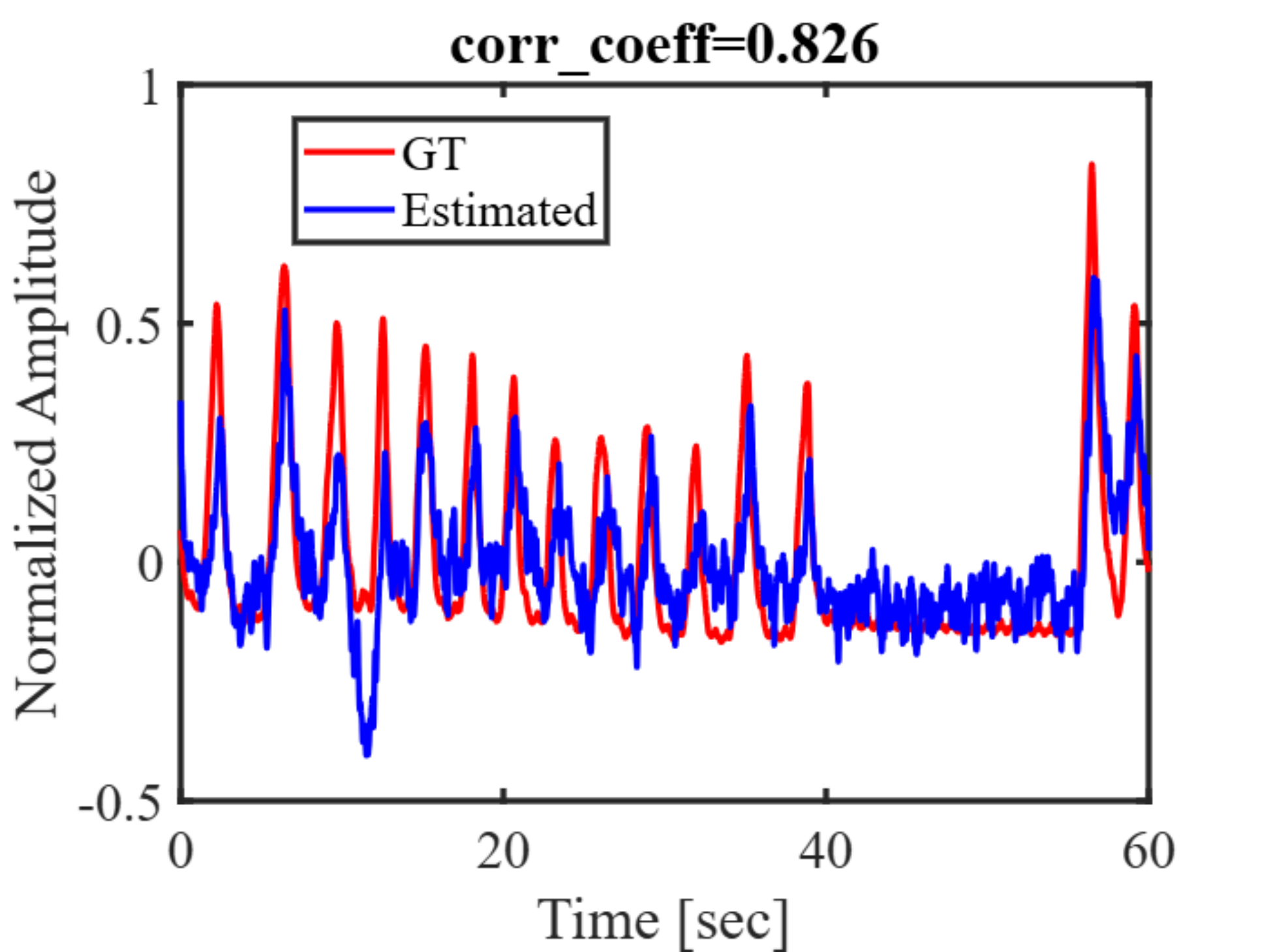}}
\end{minipage}
\hfill
\begin{minipage}[b]{0.49\linewidth}
  \centering
  \centerline{\includegraphics[width=4.8cm]{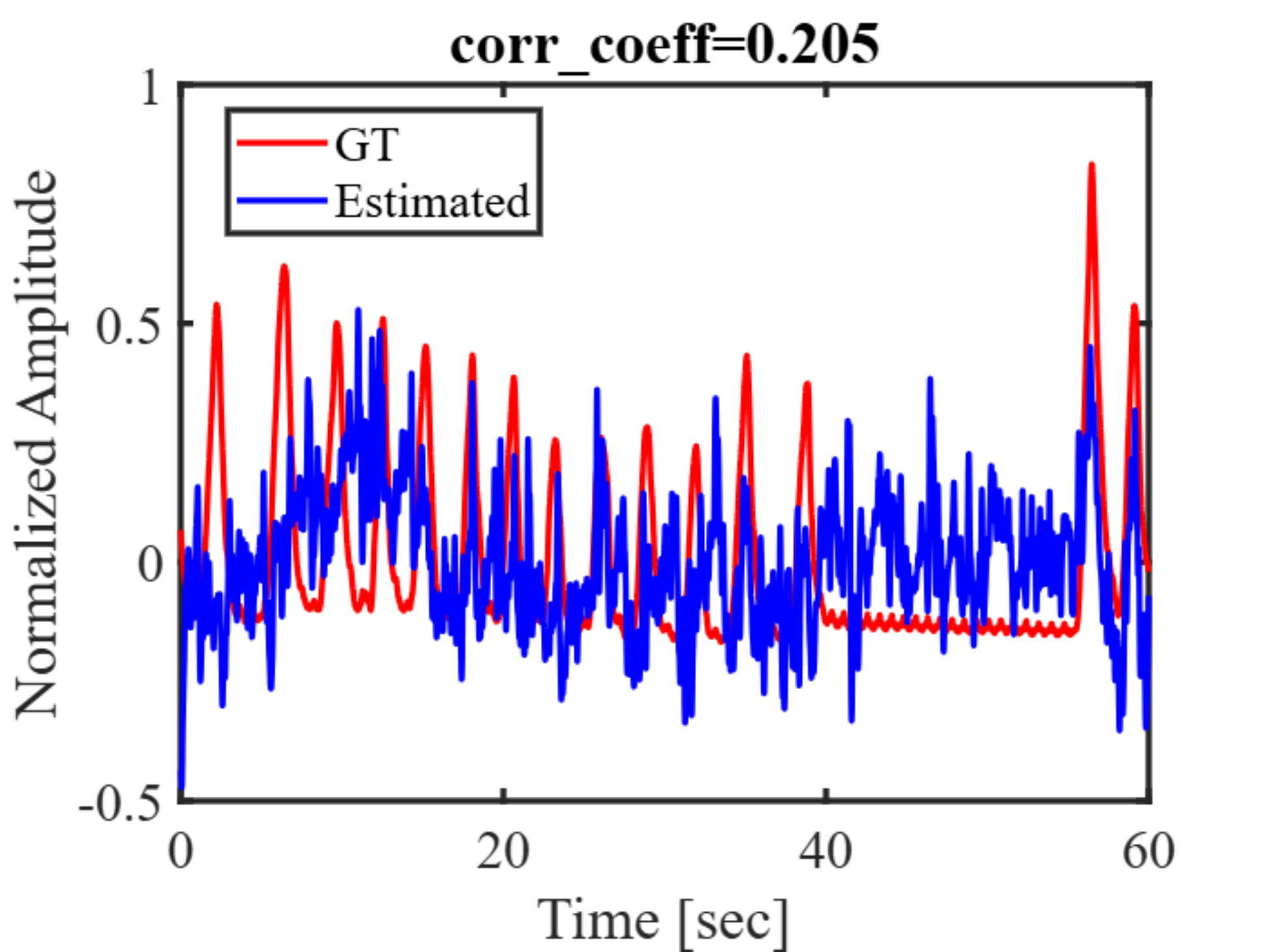}}
\end{minipage}
\caption{Example estimation from antenna 1 (left) and 57 (right).}
\label{fig:antennadiff}
\end{figure}


So far, we have estimated the breathing signal at each antenna $\hat{b}_m[i]$. However, we observe that the estimation performance varies significantly between the distributed antennas, as shown in Fig.~\ref{fig:antennadiff}.
This can be explained by Eq. (\ref{eq:delay}), the different geometry relations between the transceiver and target have different \ac{AoA}s, $\theta_{ml}[i]$, and bistatic angles $\beta_{ml}[i]$, which then affect the projection of the change of the physical path length change \cite{cherniakov2001bistatic,WiGigVital}. In addition, the multipath fading described in Eq. (\ref{eq:cfr}) also varies, leading to either constructive or destructive fading.

The $M$ preprocessed estimates $\hat{b}_m[i]$ have the same underlying breathing signal $b[i]$. This is modeled as a sensor array where each output contains a normalized estimate of the breathing signal at antenna $m$ of the sensing signal and additive noise. The resulting signal is written as follows,
\begin{equation}
\begin{aligned}
 \mb{\hat{b}}[i] = [\hat{b}_1[i] \, \hat{b}_2[i]] \, \dots \, \hat{b}_m[i]]^T ,
\end{aligned}
\end{equation}
and each antenna has a different BNR value $\Gamma_m$. Inspired by \ac{MRC} theory in MIMO systems, we combine all the antennas with the weighted sum operation. The weight of $m$-th antenna is given by  $w_m=\sqrt{\Gamma_m}$, which depends on the BNR value. Only antennas with BNR values greater than 1 are included in the combination, while those with BNR less than 1 are weighted as 0. 

\section{Experiments and Datasets}
In the following section, the measurement setup and the exact scenario for the experiments are described.

\subsection{Measurement Setup}

\textbf{The Massive MIMO Testbed: }
The RF data is collected from the Massive MIMO testbed at KU Leuven. The \ac{BS} features a Baseband Processing Unit (BBU) for high-PHY signal processing, 32 \ac{SDR} functioning as remote radio heads to handle low-PHY processing and a modular 64-antenna array, configured as eight distributed uniform linear arrays with eight antennas each.  In this experiment, we rely on the UL channel estimation at the BBU. The single-antenna \ac{UE} is also driven by SDR connected to a host computer for signal processing. Tight synchronization between the user and the \ac{BS} is achieved by connecting the local oscillators by coax cable. The \ac{UL} channel estimates are collected at a sample rate of 200 Hz. The testbed operates at a center frequency of 3.51 GHz, with 100 subcarriers spanning an 18 MHz bandwidth with a subcarrier spacing of 180 KHz. The single-antenna UE employs an omnidirectional antenna for transmission, while the receiver antennas in the BS are patch antennas. The reader is referred to \cite{datasetMagazine} for further details.

\textbf{The \ac{Mocap}:}
The ground truth is measured via Qualisys Miqus M3 Motion Capture cameras. This is a marker-based multi-camera motion capture system. 
The 3D motion capture data is collected at a sample rate of 150 Hz and is used as ground truth for the respiration estimation.

\textbf{Integration:}
The measurement is automated by simultaneously triggering the motion capture system and the Massive MIMO Testbed for data capture. This ensures time alignment between the 3D location data and the collected RF channels. The reader is referred to \cite{camad} for an in-depth description of the data logging facility.

\begin{figure}[t]
\begin{minipage}[b]{.49\linewidth}
  \centering
  \centerline{\includegraphics[width=4.2cm]{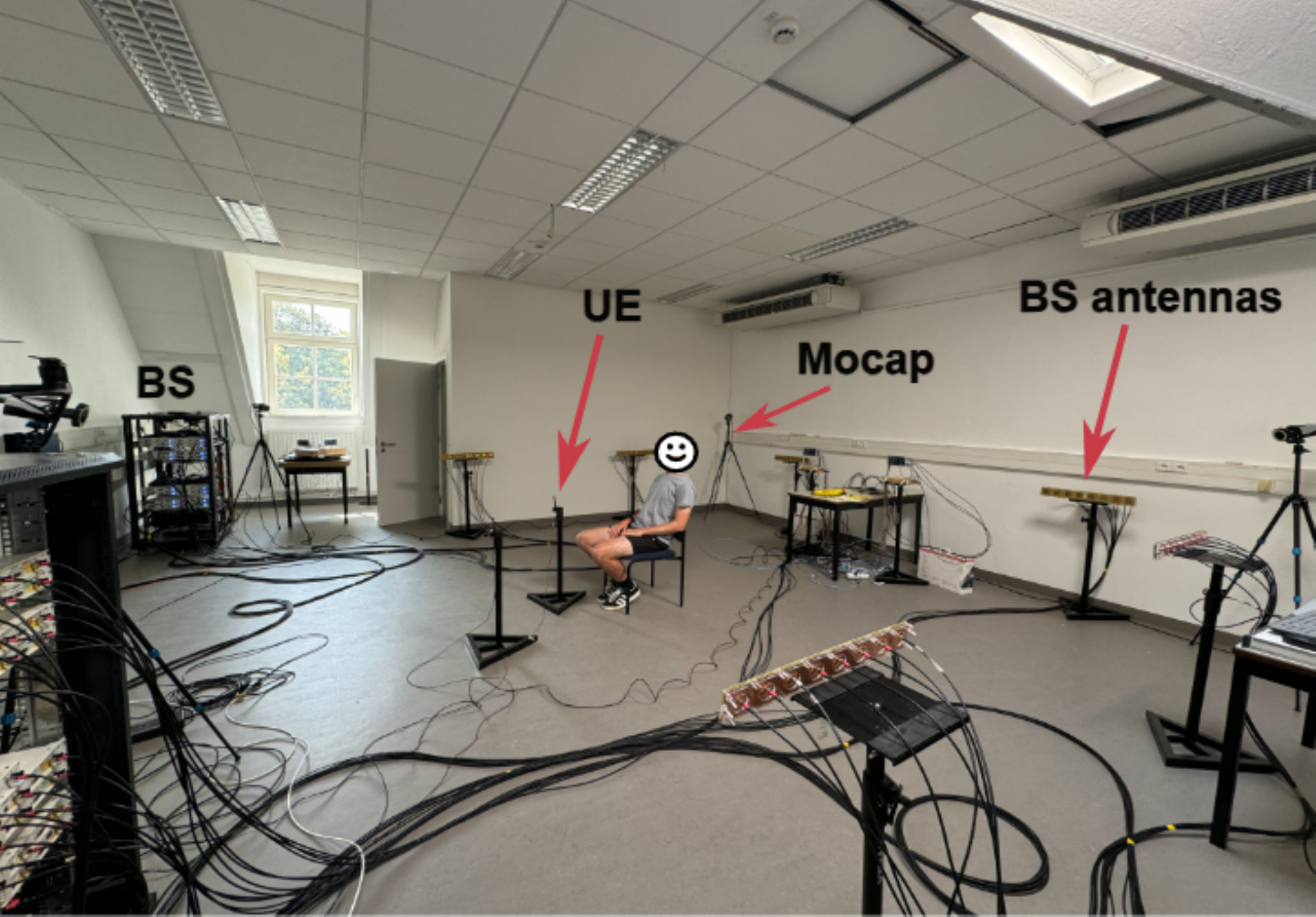}}
 \vspace{0.15cm}
  \centerline{(a) Measurement environment }\medskip
\end{minipage}
\hfill
\begin{minipage}[b]{0.49\linewidth}
  \centering
  \centerline{\includegraphics[width=4.5cm]{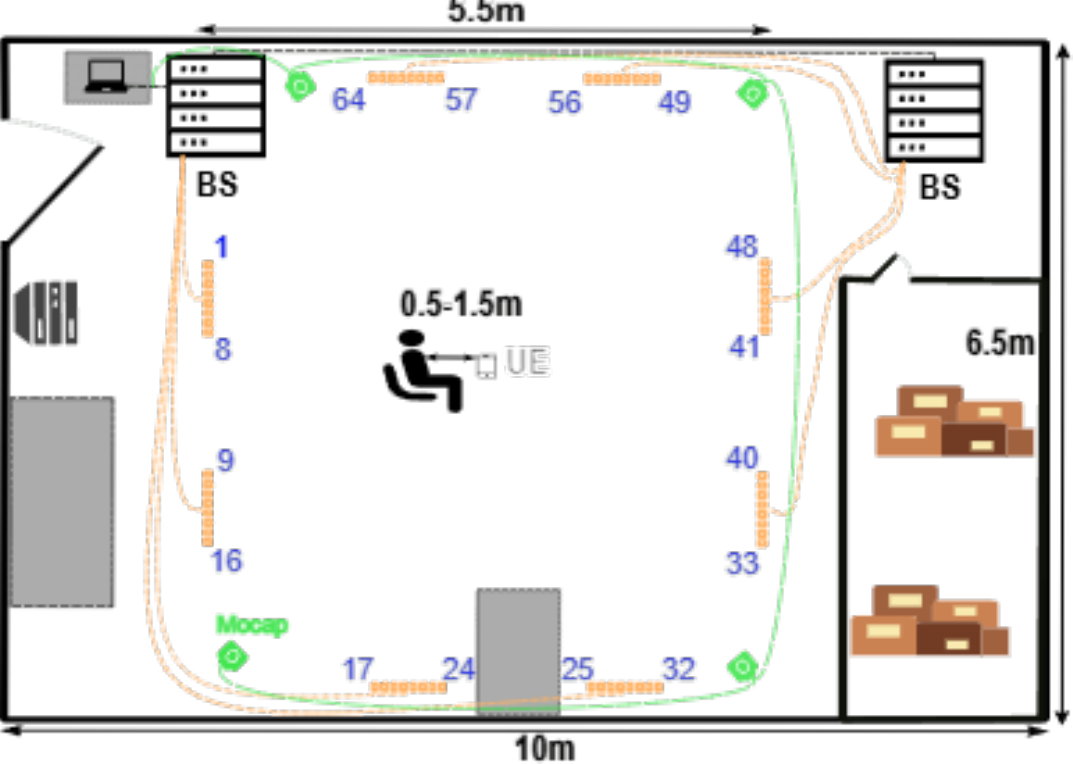}}
  \centerline{(b) Schematic}\medskip
\end{minipage}
\caption{Representation of measurement setup.}
\label{fig:setup}
\end{figure}

\textbf{Dataset:}
The \ac{BS}'s antenna arrays are positioned as depicted in Fig.~\ref{fig:setup} and the subject is seated close to the \ac{UE}. Four \ac{Mocap} cameras were deployed to record ground truth data. We collected \ac{CSI} from 10 subjects, consisting of seven males and three females, with four samples from each subject containing one minute of data each. To ensure data diversity, the \ac{UE} was moved around each subject while keeping a close but varying distance, ranging from 0.5 m to 1.5 m. We gathered 40 \ac{CSI} samples totaling 40 minutes of measurement.



\section{Numerical Performance Analysis}
\begin{figure}[tb]
    \centering
    \begin{tikzpicture}
\pgfplotstableread[col sep=comma,]{cdfPaper.csv}
\datatable
\begin{axis}[
        ylabel={Empirical CDF},
        xlabel={Correlation Coefficient},
        grid=major, 
        grid style={dashed,gray!30}, 
        major x tick style = {opacity=100},
        legend pos = north west,
                xtick align=inside,
        legend style={nodes={scale=0.6, transform shape}},
        width=\linewidth,
        height=0.8\linewidth,
         ymin=0,
         ymax=1,
         xmin=0,
         xmax=1
        ]
        \addplot+[mark repeat=32] table [x={x1}, y={y1}]{\datatable};
       \addlegendentry{SASS}
       \addplot+[mark repeat=12] table [x={x2}, y={y2}]{\datatable};
       \addlegendentry{SAMS-IDFT}
        \addplot+[mark repeat=12] table [x={x3}, y={y3} ]{\datatable};
       \addlegendentry{SAMS-DiverSense}
        \addplot+[mark repeat=4] table [x={x4}, y={y4}]{\datatable};
        \addlegendentry{MAMS-WAC}

        \end{axis}
\end{tikzpicture}
    \caption{Correlation of different estimation methods with ground truth}
    \label{fig:cdf}
\end{figure}
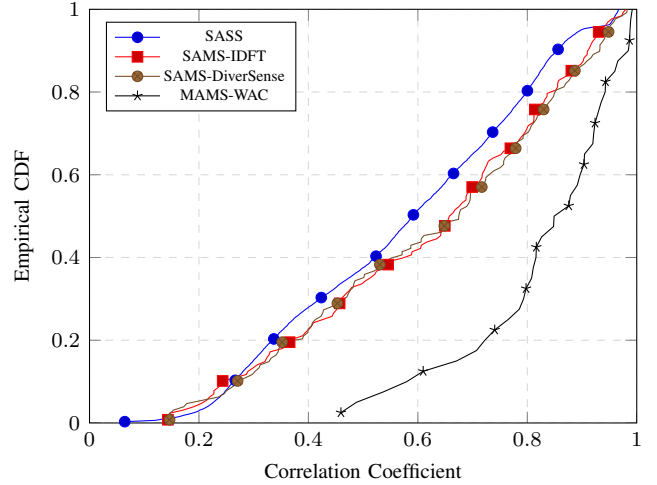

The following section contains the results of our experiments.
To demonstrate the capabilities of Massive MIMO OFDM sensing, we consider three distinct processing methods: 1) Single-Antenna and Single-Subcarrier (SASS),  2) Single-Antenna and Multiple-Subcarriers (SAMS), and 3) Multiple-Antennas and Multiple-Subcarriers (MAMS). They are compared shown in Fig. \ref{fig:cdf}.  


\textbf{Single-Antenna and Single-Subcarrier:}
We start by utilizing a single subcarrier for respiration estimation. The cumulative probability curve shows that only 5.2\% of subcarriers have a correlation value over 0.9. 50\% of subcarriers have a correlation value less than 0.6. This suggests that the estimation from a single subcarrier provides poor accuracy.

\textbf{Single-Antenna and Multiple-Subcarriers:}
This utilizes the combining method described in Sec.~\ref{sams}
when expanding the estimation to multiple subcarriers on a single antenna via \ac{IDFT} and DiverSense. Both methods show a comparable improvement, 
while the \ac{IDFT} performance is slightly better. This indicates that both methods effectively capitalize on the information across multiple subcarriers to enhance estimation accuracy. However, the improvement is very limited due to the narrow 18 MHz bandwidth.

\begin{figure}[t]
\begin{minipage}[b]{.49\linewidth}
  \centering
  \centerline{\includegraphics[width=4.8cm]{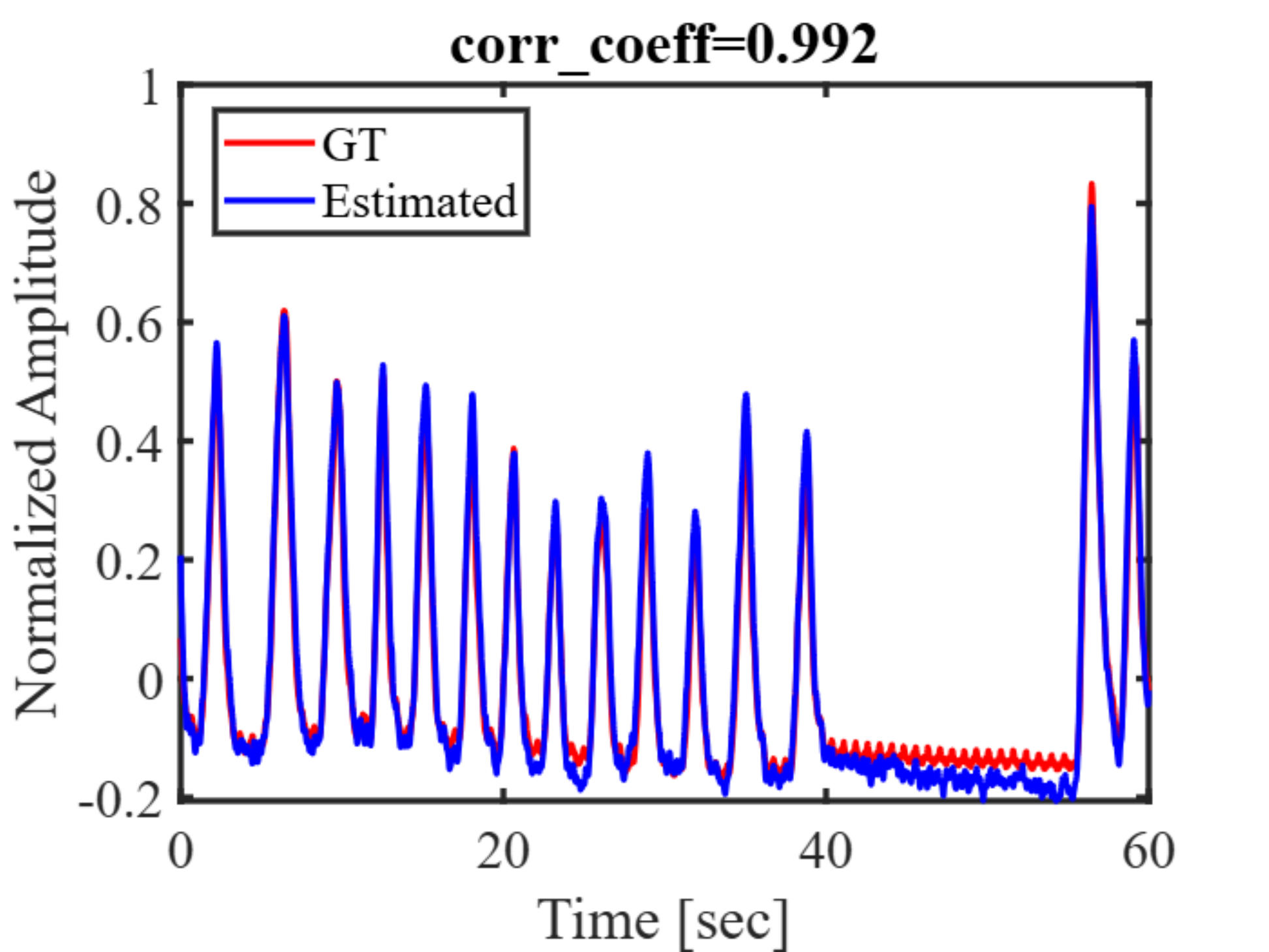}}
  \centerline{(a) subject 1 }\medskip
\end{minipage}
\hfill
\begin{minipage}[b]{0.49\linewidth}
  \centering
  \centerline{\includegraphics[width=4.8cm]{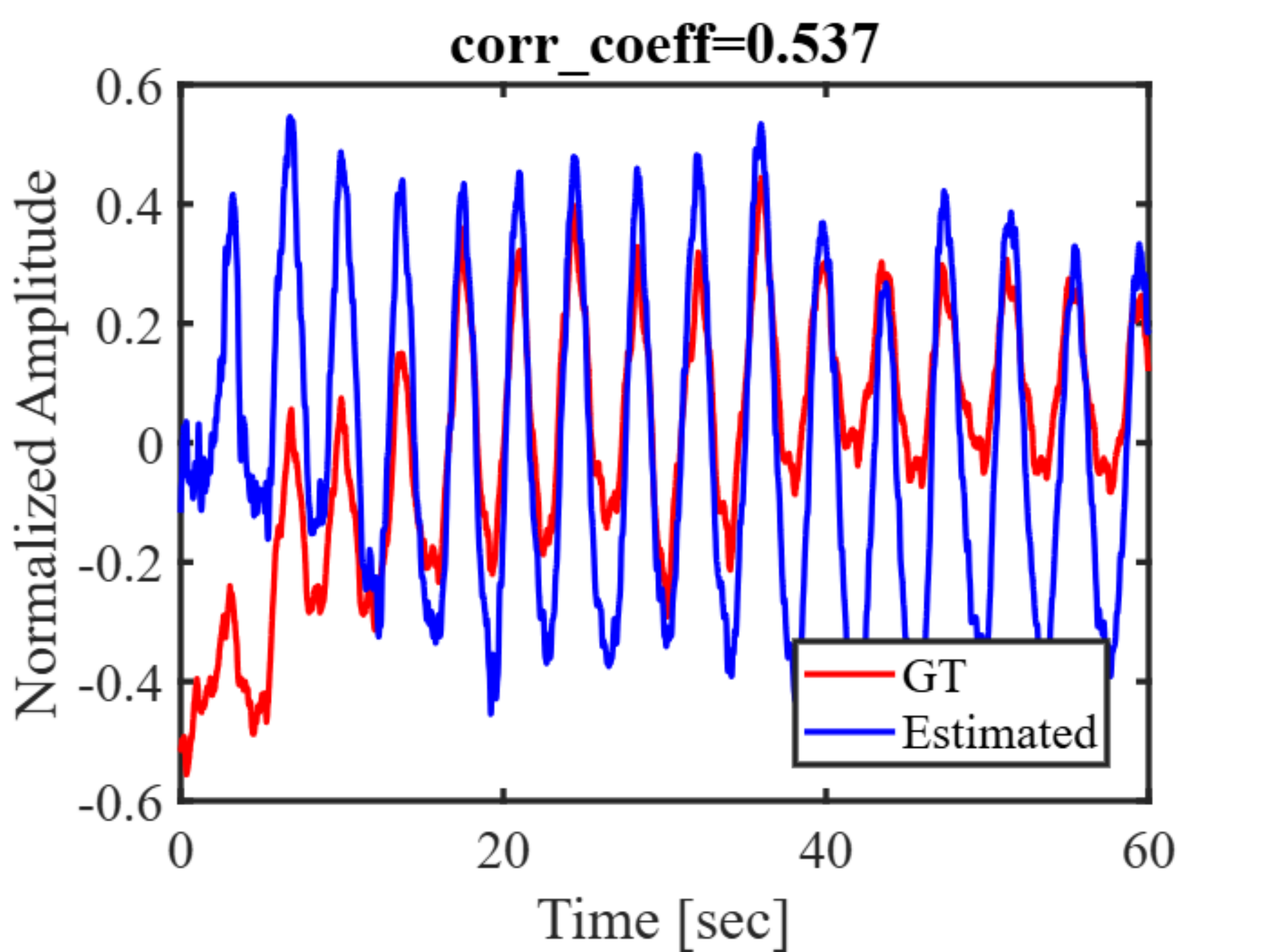}}
  \centerline{(b) subject 9}\medskip
\end{minipage}
\caption{Example estimation from subject (a) 1 and (b) 9 via MAMS-WAC.}
\label{fig:example}
\end{figure}

\textbf{Multiple-Antennas and Multiple-Subcarriers:}
Finally, we assess the performance when leveraging data from multiple subcarriers and antennas. Since \ac{IDFT} shows better robustness, we use it in \ac{WAC}.  We show two example results from two subjects in Fig. \ref{fig:example}, with a correlation of 0.992 and 0.537 with corresponding ground truth. Notice that the bad correlation value from subject 17 is because of the 'noisy' ground truth. The CDF curve surpasses the single antenna methods around the 0.2 correlation value on average. It maintains a higher probability of achieving a closer correlation with the ground truth across all observed values, which demonstrates the best performance. This indicates a substantial improvement in estimation performance, highlighting the advantages of integrating signals over multiple antennas and subcarriers for robust and accurate breathing estimation.

\section{Conclusion}
This paper used a Massive MIMO OFDM testbed to conduct passive respiration sensing. We explored signal processing techniques such as IDFT, DiverSense, and \ac{WAC} to improve the accuracy and robustness of respiration sensing. The method achieved perfect bpm estimation and an average correlation of 0.8 with the ground truth, even with a narrow bandwidth of 18 MHz. This research confirms the potential of massive MIMO systems in health monitoring applications, verifying the feasibility of utilizing communication infrastructure for dual-functional purposes.


\balance
\bibliographystyle{ieeetr} 
\bibliography{main} 


\end{document}